\documentclass[12pt]{article}
\usepackage{epsfig}
\date{\today}
 \normalsize

\newcommand{\bmat}{\left(\begin{array}}
\newcommand{\emat}{\end{array}\right)}
\newcommand{\be}{\begin{equation}}
\newcommand{\ee}{\end{equation}}
\newcommand{\bea}{\begin{eqnarray}}
\newcommand{\eea}{\end{eqnarray}}

\def\nc{noncommutative }

\def    \be            {\begin{equation}}
\def    \ee            {\end{equation}}
\def    \bea           {\begin{eqnarray}}
\def    \eea           {\end{eqnarray}}

\begin{document}
\renewcommand{\thefootnote}{\fnsymbol{footnote}}
\vspace{.3cm}

\title{\Large\bf The Van der Waals interactions and the photoelectric effect in \nc quantum mechanics }

\author
{ \it \bf   K. Li$^{1}$\thanks{kangli@hztc.edu.cn} and N.
Chamoun$^{2,3}$\thanks{nchamoun@ictp.it}\\ \small $^1$ Department
of Physics, Hangzhou Normal University, Hangzhou 310036, China
\\ \small$^2$ The Abdus Salam ICTP, P.O. Box 586, 34100
Trieste, Italy. \\
\small$^3$ Physics Department, HIAST, P.O.Box 31983, Damascus,
Syria. }
\date{}
\maketitle

\begin{center}
\small{\bf Abstract}\\[3mm]
\end{center}
We calculate the long-range Vanderwaals force and the
photoelectric cross section in a \nc set up. While we argue that
non-commutativity effects could not be discerned for the
Vanderwaals interactions, the result for the photoelectric effect
shows deviation from the usual commutative one, which in principle
can be used to put bounds on the space-space non-commutativity
parameter.

 \vspace{1.1cm}{\bf Keywords}: \nc space, Vanderwaals forces,
 photoelectric effect
\begin{minipage}[h]{14.0cm}
\end{minipage}
\vskip 0.3cm \hrule \vskip 0.5cm
\section{{\large \bf Introduction}}
A large amount of research work has been devoted to the study of
physics on \nc space-times (for a review see, e.g.,
\cite{review}). This was motivated by the discovery in string
theory that the low energy effective theory of D-brane in the
background of NS-NS $B$ field lives on noncommutative space
\cite{CDS}-\cite{SW}.

In the \nc space, the coordinate and momentum operators verify
\be\label{Eq:nmr2} ~[\hat{x}_{i},\hat{x}_{j}]=i\theta_{ij},~~~
[\hat{p}_{i},\hat{p}_{j}]=0,~~~[\hat{x}_{i},\hat{p}_{j}]=i
\hbar\delta_{ij}, \ee where $\hat{x}_i$ and $\hat{p}_i$ are the
coordinate and momentum operators, and where we assume the time
coordinate is commutative. The parameter $\{\theta_{ij}\}$ is an
antisymmetric matrix representing the non-commutativity of the
space, and is of dimension $(length)^2$. In many proposals to test
the hypothetical spacetime noncommutativity, one does not need the
exact quantum field theory, but only its quantum mechanical
approximation, and many simple quantum mechanics (QM) problems
were treated on \nc spaces. For instance, the Hydrogen atom
spectrum and the Lamb shift in \nc quantum electrodynamics (QED)
were first treated in \cite{CST} (look also at
\cite{Ho,JabbComment}). The \nc version of QED has been examined
in \cite{ncqed}. A method for formulating the non-abelian \nc
field theories has been discussed in \cite{nonabelian} and using
these ideas \nc version of the standard model (SM) has been
proposed \cite{ncsm}. In this \nc version, there are several new
features and interactions, like triple gauge boson vertices,
 that appear and some of the related phenomenological aspects of these
 have been investigated \cite{ncsmpheno}. Also, the
questions of $t \to b W$ scattering in \nc SM and the PCT theorem
in \nc fields theory were treated in \cite{namit}. Fractional
quantum Hall effect could be obtained using \nc rank 1
Chern-Simons theory \cite{poly01}, and \nc massive Thirring model
was treated in\cite{MNR} . The QM problems could be treated
equally in \nc phase space \cite{djemai,dulat}, where momenta are
also \nc enabling, thus, to incorporate an additional background
magnetic field and to maintain Bose-Einstein statistics
\cite{douglas,Likang}.

However, although \nc QM has been extensively studied, there are
still some problems which were not treated in the \nc set up. To
our knowledge, the phenomenological implications of
noncommutativity on the Vanderwaals interaction between molecules
and on the photoelectric effect were not examined, and the subject
of this letter is just to present such an analysis.

For the Vanderwaals forces, we find that the \nc effects can not
be determined since they are far smaller than the next term to the
dominant perturbative one proprtional to $1/R^6$ where $R$ is the
interatomic distance. As for the photoelectric effect, we find
that the non-commutativity introduces a phase into the
corresponding transition matrix element proportional to the
parameter $\theta$ of space-space non-commutativity. Thus, the
photoelectric cross section gets multiplied by a factor
proportional to $\theta^2$, which can be used to put bounds on
this parameter.

\section{Analysis}
To start, we note that the action for field theories on \nc spaces
is obtained from the usual commutative action by replacing each
usual product of fields ($f \cdot g$) by the star-product:

\be\label{star} (f*g)(x)=exp({i\over
2}\theta_{\mu\nu}\partial_{x_{\mu}}\partial_{y_{\nu}})
f(x)g(y)|_{x=y}\ , \ee where $f$ and $g$ are two arbitrary
infinitely differentiable functions on $R^{3+1}$.

Alternatively, one can change the \nc problems into problems of
familiar commutative spaces using the new \nc variables ($\hat{x}$
and $\hat{p}$) defined in terms of the commutative ones ($x$ and
$p$) by: \bea \hat{x}_i &=& x_i - \frac{1}{2\hbar}\theta_{ij}p_j
\nonumber \\ \hat{p}_i &=& p_i \eea and the Hamiltonian
corresponding to the \nc problem is obtained by replacing ($x$ and
$p$) in the commutative Hamiltonian by ($\hat{x}$ and $\hat{p}$).
In three dimensions, we can define the vector {\mbox{\boldmath
{$\theta$}}} by $\theta_{ij}=\frac{1}{2} \epsilon_{ijk}\theta^k$,
so that the \nc coordinate becomes \bea \hat{{\mbox{\boldmath
{$x$}}}}&=& {\mbox{\boldmath {$x$}}}-\frac{1}{4\hbar}
{\mbox{\boldmath {$p$}}}\times {\mbox{\boldmath {$\theta$}}}\eea

\subsection{Vanderwaals forces}
We discuss here the known problem of the long-range interaction
between two Hydrogen atoms in their ground states. We assume the
nuclei of the two atoms are fixed in space a distance $R$ apart
and we choose the $z$ axis parallel to the line between them. Let
${\mbox{\boldmath {$r$}}}_1$ be the vector displacement of the
electron $1$ from the nucleus $A$ and ${\mbox{\boldmath {$r$}}}_2$
be the vector displacement of the electron $2$ from the nucleus
$B$, then the `ordinary' hamiltonian for the two electrons can
be written \bea \label{cVanHam} H^c &=& H^0 + H'^c \\
H^0 &=& -\frac{\hbar^2}{2m}\left({\mbox{\boldmath {$\nabla$}}}_1^2
+ {\mbox{\boldmath {$\nabla$}}}_2^2\right)
-\frac{e^2}{r_1}-\frac{e^2}{r_2} \\
H'^c&=&\frac{e^2}{R}+\frac{e^2}{r_{12}}-\frac{e^2}{r_{1B}}-\frac{e^2}{r_{2A}}\eea
Expanding $H'^c$ in powers of $r/R$, where $r\sim r_1 \sim r_2$ of
order of $a_0$ the Bohr radius, we find \cite{schiff} \bea H'^c&=&
H'^c_3+{\cal{O}}\left(\frac{r^3}{R^4}\right)\\ H'^c_3 &=&
\frac{e^2}{R^3}\left(x_1x_2+y_1y_2-2z_1z_2 \right)\eea The last
term $H'^c_3$ represents the interaction energy of two electric
dipoles that correspond to the instantaneous configurations of the
two atoms, while the higher order terms denote the
quadrupole-dipole, quadrupole-quadrupole and higher order
interactions. The unperturbed hamiltonian $H^0$ has the solution
\bea u\left({\mbox{\boldmath {$r$}}}_1,{\mbox{\boldmath
{$r$}}}_2\right)&=&u_{100}\left({\mbox{\boldmath
{$r$}}}_1\right)u_{100}\left({\mbox{\boldmath {$r$}}}_2\right)\eea
for two interacting hydrogen atoms in their ground state.

In the \nc space, and where we assume $R$ as a fixed parameter, we
then have \bea H&\equiv&
H^0\left(\hat{r},\hat{p}\right)+H'^c\left(\hat{r},\hat{p}\right)\\&=&
H^0\left(r,p\right)+H'\left(r,p\right)\\ H'\left(r,p\right)&=& H'^c + H'^{NC}\\
H'^{NC}&=& H'^{NC}_1 + H'^{NC}_2 \eea where $H'^{NC}_1$
($H'^{NC}_2$) comes from the non-commutativity effects on $H^0$
($H'^c$)\bea
H'^{NC}_1&=&H_1^{\theta}+{\cal{O}}\left(\frac{\theta^2}{r^5}\right)\\
H'^{NC}_2&=& H_2^{\theta}+{\cal{O}}\left(\frac{\theta
r}{R^4},\frac{\theta^2}{R^3 r^2}\right) \eea with \bea
H_1^{\theta}&=& -\frac{e^2}{4\hbar}{\mbox{\boldmath
{$\theta$}}}\cdot \left( \frac{{\mbox{\boldmath
{$L$}}}_1}{r^3_1}+\frac{{\mbox{\boldmath {$L$}}}_2}{r^3_2}
\right)\\H_2^{\theta}&=& -\frac{e^2}{4\hbar R^3}\left[
\theta^x\left( y_1p_{2z}+y_2p_{1z}+2z_1p_{2y}+2z_2p_{1y}
\right)\right. \nonumber
\\ &&\left. + \theta^z\left(
x_1p_{2y}+x_2p_{1y}-y_1p_{2x}-y_2p_{1x} \right) \right] \eea Here
we assumed, without loss of generality, that the vector
{\mbox{\boldmath {$\theta$}}} has components along the $z$ and $x$
axes only.

It is clear that the expectation value of the leading terms
($H'^c_3,H^\theta_1,H^\theta_2$) for the state
$u\left({\mbox{\boldmath {$r$}}}_1,{\mbox{\boldmath
{$r$}}}_2\right)$ is zero. This is because $u_0$ is an even
function of ${\mbox{\boldmath {$r$}}}_1$ and ${\mbox{\boldmath
{$r$}}}_2$ while $H'^c_3,H^\theta_2$ are odd functions of
${\mbox{\boldmath {$r$}}}_1$ and ${\mbox{\boldmath {$r$}}}_2$
separately, and $H^\theta_1$ applied to $u_0$ would give zero. One
can check, neglecting the terms proportional to $\theta^2$, that
all the neglected higher order in the perturbation $H'$ have zero
expectation value for $u_0$. Thus the leading term in the
interaction energy is the second-order perturbation of the
dipole-dipole which is proportional to $(H'^c_3)^2$ and hence
varies like $1/R^6$. This is the well known nature of the
Vanderwaals force.

The non-commutativity would cause a shift in the interaction
energy estimated by: \bea \Delta E = \sum_{n\neq
0}\frac{|<0|H'^{NC}|n>|^2}{E_n-E_0} + 2 \sum_{n\neq
0}\frac{<0|H'^c|n><n|H'^{NC}|0>}{E_n-E_0}
 \eea
We see here that the perturbation $H^\theta_1$ does not contribute
to the matrix element $<0|H'^{NC}|n>$, since $L|0>=0$, and thus
does not lead to any energy shift in the Vanderwaals interaction.
We would like to mention here that in the case of excited states,
the perturbation $H^\theta_1$ can be perceived only for very
`dilute' gas. The reason of this is as follows. Since $H^\theta_1
\sim \frac{\theta}{r^3}$, then keeping $H^\theta_1$ while dropping
the next term to $H'^c_3$ in the expansion of $H'^c$ which is
proportional to $\frac{r^3}{R^4}$ implies $\frac{r^4}{R^4} <
\frac{\theta}{r^2}$. Equally dropping the next term to
$H^\theta_1$ in the expansion of $H'^{NC}_1$ proportional to
$\frac{\theta^2}{r^5}$ while keeping $H'^c_3 \sim \frac{r^2}{R^3}$
implies $\frac{\theta}{r^2} <
\left(\frac{r}{R}\right)^{\frac{3}{2}}$. Thus we should have \bea
\left(\frac{r}{R}\right)^4 < \frac{\theta}{r^2} <
\left(\frac{r}{R}\right)^{\frac{3}{2}}\eea The bound $\theta <
10^{-8}$ GeV$^{-2}$ \cite{CST} (look at \cite{IMR} for tighter
bounds) with the ordinary value of $r \sim a_0 \sim 10^{-10} m$,
would give a bound $\frac{\theta}{r^2} < 10^{-20}$. Hence, if \nc
effects for the term $H^\theta_1$ would be tangible in the
Vanderwaals interaction between two `excited' atoms, the gas of
molecules should be `dilute' enough satisfying $\frac{r}{R} <
10^{-5}$.

As to the term $H^\theta_2$, we should determine to which order
${\cal{O}}\left(\frac{r^{n-1}}{R^n}\right)$ we should expand
$H'^c$ so that not to drop $H^\theta_2$. For the value
$\frac{r}{R} \sim 10^{-5}$ and the bound $\theta < 10^{-8}$
GeV$^{-2}$, we see that we should expand $H'^c$ till the order
$n=7$, while for normal gases $\frac{r}{R}\sim 10^{-1}$, we see
that we can not feel $H^\theta_2$ unless we expand $H'^c$ till
$n=23$. We conclude that non-commutativity effects can not be
determined for the Vanderwaals interactions.

\subsection{The photoelectric effect}
We now consider the photoelectric effect-that is, the ejection of
an electron when an atom is placed in the radiation field. The
basic process is considered to be the transition from an atomic
(bound) state to a continuum state ($E>0$). For the final state
$|{\mbox{\boldmath {$f$}}}^0>$, we must use a positive energy
eigenstate of the Coulomb Hamiltonian \[H^0 =
\frac{p^2}{2m}-\frac{e^2}{r}\] However, if the ejected electron is
not too slow, then one can ignore the pull of the proton on it
and, with negligible error, approximate the continuum state with a
plane wave state $|{\mbox{\boldmath {$p$}}}_f>$ with momentum
${\mbox{\boldmath {$p$}}}_f$:
\[|{\mbox{\boldmath {$f$}}}^0>
 = |{\mbox{\boldmath {$p$}}}_f> + \cdots\] an approximation which, moreover, assumes that
$|{\mbox{\boldmath {$p$}}}_f>$ is dominating the higher order
terms in $|{\mbox{\boldmath {$f$}}}^0>$ when evaluating matrix
elements\cite{bethe}.

We consider now the Hydrogen atom in its ground state $u_{100}$ on
which is incident the electromagnetic wave \bea {\mbox{\boldmath
{$A$}}}\left({\mbox{\boldmath {$r$}}},t\right)&=& {\mbox{\boldmath
{$A$}}}_0 e^{i\left({\mbox{\boldmath {$k$}}}\cdot{\mbox{\boldmath
{$r$}}}-wt\right)}\eea We would like to calculate the rate for the
process of liberating a bound electron using the Fermi's golden
rule: \bea R_{i \to f} &=rate \, of\, transition\, i\to f=&
\frac{2\pi}{\hbar}|<f^0|H'|i^0>|^2 \delta\left(E_f^0-E_i^0-\hbar
w\right)\eea where, as we said, the final state $|f^0>$ is the
plane wave $|{\mbox{\boldmath {$p$}}}_f>$ while the initial state
$|i^0>$ is the ground state $|100>$.

The `ordinary' perturbation, in the Coulomb gauge
${\mbox{\boldmath {$\nabla$}}}\cdot {\mbox{\boldmath {$A$}}} =0$,
is \cite{shankar} \bea H'^c(t)&=\frac{e}{2mc}\,
 e^{i{\mbox{\boldmath
{$k$}}}\cdot{\mbox{\boldmath {$r$}}}} e^{-iwt} {\mbox{\boldmath
{$A$}}}_0 \cdot {\mbox{\boldmath {$p$}}}=& H^{1c} e^{-iwt}\eea

Now, we introduce the non-commutativity so to obtain for the
perturbation:\bea H^1&=& \frac{e}{2mc}\,
 e^{i{\mbox{\boldmath
{$k$}}}\cdot \left({\mbox{\boldmath
{$r$}}}-\frac{1}{4\hbar}{\mbox{\boldmath {$p$}}} \times
{\mbox{\boldmath {$\theta$}}}\right)} {\mbox{\boldmath {$A$}}}_0
\cdot {\mbox{\boldmath {$p$}}}\eea  We can proceed now to evaluate
the transition matrix element in the coordinate basis \bea
H^1_{fi} &=& N \int e^{-i {\mbox{\boldmath {$p$}}}_f \cdot
{\mbox{\boldmath {$r$}}}/\hbar} e^{i{\mbox{\boldmath {$k$}}}\cdot
\left({\mbox{\boldmath {$r$}}}-\frac{1}{4\hbar}{\mbox{\boldmath
{$p$}}} \times {\mbox{\boldmath {$\theta$}}}\right)}
{\mbox{\boldmath {$A$}}}_0 \cdot \left(-i \hbar {\mbox{\boldmath
{$\nabla$}}} \right) e^{-r/a_0} d^3{\mbox{\boldmath {$r$}}}
\\N&=& \frac{e}{2mc}\frac{1}{\left(2\pi \hbar\right)^{3/2}} \left(\frac{1}{\pi a_0^3}\right)^
{1/2}\eea where $a_0$ is the Bohr radius for the ground state
Hydrogen atom.

Notice here, that we take the factor $e^{i{\mbox{\boldmath
{$k$}}}\cdot{\mbox{\boldmath {$r$}}}}$ exactly into account, i.e.
we do not assume the electric dipole approximation. Integrating by
part and using ${\mbox{\boldmath {$A$}}}_0 \cdot {\mbox{\boldmath
{$k$}}}=0$, we get up to first order in $\theta$: \bea \label{Hfi}
H^1_{fi} &=& N
\frac{\left(8\pi/a_0\right)^2}{\left[\left(1/a_0\right)^2+\left({\mbox{\boldmath
{$p$}}}_f/\hbar - {\mbox{\boldmath {$k$}}}\right)^2\right]^2}
{\mbox{\boldmath {$A$}}}_0 \cdot {\mbox{\boldmath {$p$}}}_f
\left[1-\frac{i}{4\hbar}  \left( {\mbox{\boldmath
{$\theta$}}}\times {\mbox{\boldmath {$k$}}}\right) \cdot
{\mbox{\boldmath {$p$}}}_f\right]\eea

Like the ordinary case, we see that the rate depends on the
magnitude of the applied field ${\mbox{\boldmath {$A$}}}_0$ and on
its angle with the outgoing momentum, and also on the frequency of
the radiation. The electron likes to come parallel to
${\mbox{\boldmath {$A$}}}_0$, but is also biased towards
${\mbox{\boldmath {$k$}}}$. However, ejecting the electron
parallel to ${\mbox{\boldmath {$k$}}}$ minimizes the denominator
in equation (\ref{Hfi}), but nullifies the \nc effect. The \nc
effect consists of introducing an imaginary part into the
transition matrix element. If our detector counts how many
electrons come in the cone of solid state $d\Omega$, then we can
associate with the atom a photoelectric differential cross section
\bea \frac{d\sigma}{d\Omega} &=& \frac {8 \pi c}{|{\mbox{\boldmath
{$A$}}}_0|^2w^2} \cdot \hbar w \cdot R_{i \to d\Omega}\eea with
the final result: \bea
\left(\frac{d\sigma}{d\Omega}\right)^{NC}&=&\left(\frac{d\sigma}{d\Omega}\right)^{c}\left[1+\frac{1}{16\hbar^2}
\left( \left( {\mbox{\boldmath {$\theta$}}}\times {\mbox{\boldmath
{$k$}}}\right) \cdot {\mbox{\boldmath {$p$}}}_f\right)^2\right]
 \eea
This formula can be used to put bounds on the parameter $\theta$.
However, since the correction is in $\theta^2$ (quadratic), the
deviation between the experimental data and the ordinary cross
section $\left(\frac{d\sigma}{d\Omega}\right)^{c}$ will not lead
to stronger bounds on $\theta$ than the Lamb shift bound
$\left(\theta < 10^{-8}GeV^{-2}\right)$ where it enters linearly
\cite{CST}.

\section{{\large \bf Conclusions}}
We have studied the problems of the \nc Vandrwaals interactions
and photoelectric effect. For the Vanderwaals force, the \nc
effect can not be determined experimentally since the errors
coming from neglecting higher orders in the `ordinary'
perturbative expansion are far larger than the \nc corrections.
For the photoelectric effect, the cross section is multiplied by a
factor proportional to $(1+\theta^2)$ allowing to put bounds on
the \nc parameter $\theta$.

\section*{{\large \bf Acknowledgements}}
Major part of this work was done within the Associate Scheme of
ICTP. We thank Namit Mahajan for useful discussions.

\end{document}